\documentclass[twocolumn,showpacs,amsmath,amssymb,pra,floatfix]{revtex4}

\usepackage{times}
\usepackage{bm} 
\usepackage{epsf} 
\usepackage[english]{babel} 
\usepackage{graphicx} 
\usepackage{amsmath}

\newcommand{\hmol}[0]{H$_2$}% 
\newcommand{\hscal}[0]{H$_{\rm scal}$}%
\newcommand{\Rn}[0]{R_{\rm nu}}%

\newcommand{\braket}[2]{\langle \, #1 \, | \, #2 \, \rangle}

\newcommand{\opp}[1]{\hat{#1}}

\newcommand{\diff}[1]{\textrm{d}#1}

\begin{document}

\bibliographystyle{apsrev} 
%\bibliographystyle{unsrt} 

% 
%%%%%%%%%%%%%%%%%%%%%%%%%%%%%%%%%%%%%%%%%%%%%%%%%%%%%%%%%%%%%%%%%%% 
% 
 
\title{Full two-electron calculations of antiproton collisions with molecular
  hydrogen}       
 
\author{Armin L\"uhr} 
\author{Alejandro Saenz} 
 
\affiliation 
{Institut f\"ur Physik,  
AG Moderne Optik, Humboldt-Universit\"at zu Berlin, Newtonstr. 15, 
D-12489 Berlin, Germany}  
%\ead{\mailto{Armin.Luehr@physik.hu-berlin.de}} 

\date{\today} 
 
             
\begin{abstract}\label{txt:abstract} 
Total cross sections for single ionization and excitation of molecular
hydrogen by antiproton impact are presented  over a wide range of impact energy
from 1 keV to 6.5 MeV. 
A nonpertubative time-dependent close-coupling method
is applied to fully treat the correlated dynamics of the electrons. Good
agreement is obtained between the present calculations and experimental
measurements of single-ionization cross sections at high energies,
whereas some discrepancies with the experiment are found around the
maximum. 
The importance of the molecular geometry and a full two-electron
description is demonstrated.
The present findings provide benchmark results 
which might be useful for the development of molecular models.
%for calculations 
%existing data which were calculated 
%using 
%one-electron 
%model potentials. 
%A comparison with two simple models yields their range of applicability. 
%of these models as well as 
%The importance of molecular, orientation, and electron-electron correlation
%effects is discussed.  
\end{abstract}

\maketitle

%%%%%%%%%%%%%%%%%%%%%%%%%%%%%%%%%%%%%%%%%%%%%%%%%%%%%%%%%%%%%%%%%%%%%%%%%% 

%\section{Introduction} 
%\label{sec:introduction}

A central point of atomic and molecular physics is the description of
charged particles moving in a Coulomb field.
One of the simplest and most basic systems which provides an insight into
% allows for the study of
dynamic processes of charged particles is the collision of antiprotons with
atoms. The heavy mass of the antiproton allows, first, for a 
semiclassical theoretical approach and, second, for the investigation of
``slow'' ionizing collisions. 
In contrast to positively charged projectiles, for antiprotons there is no
complication from charge transfer.

Further attention is drawn to this topic due to the upcoming FAIR
\cite{anti:fair} facility
%Facility for Antiproton and Ion Research (FAIR). 
with the international collaborations FLAIR \cite{anti:flai} and SPARC 
\cite{anti:spar}, both intending to investigate antiproton driven processes
and  even kinematically complete collision experiments.
However, the design of FLAIR already requires a reliable knowledge of
low-energy antiproton cross sections of residual gases as, e.g.,  molecular
hydrogen.  
These %relatively new 
experimental efforts complement the recent intensive
studies on antihydrogen at CERN aiming to test the CPT
invariance and to disclose the nature of antimatter gravity.

Over the last decades a remarkable progress in the understanding of
interactions between antiprotons and atoms has been achieved (cf.\
\cite{anti:knud08} and Refs.\ therein).  
The theoretical description concentrated mainly on hydrogen and helium atoms. 
It was relatively easy to establish a full treatment of the former target
which only consists of one electron and one nucleus. 
%due to the lack of electron-electron correlation 
For helium atoms, on the other hand, the {\em dynamic}
electron-electron correlation effects turned out to be decisive; requiring much
larger efforts for their correct description.
Due to the above mentioned favorable properties, antiproton collisions on
helium atoms became a benchmark system for studying electron correlation in
atoms stimulating a large number of calculations which employed various
theoretical methods.
% effects requiring  much larger efforts for their correct description.
During the last ten years close-coupling calculations using either a
\emph{spectral} or \emph{spatial} expansion of the two-electron wavefunction
\cite{anti:lee00,anti:igar04,anti:schu03,anti:fost08} yielded the most precise
results where the latter usually take advantage of large-scale computing
facilities. 
% Close-coupling calculations using either a \emph{spectral} or \emph{spatial}
% expansion of the two-electron wavefunction
% \cite{anti:lee00,anti:igar04,anti:schu03,anti:fost08} have been leading the
% competition for the highest precision during the last ten years where the
% latter usually take advantage of large-scale computing facilities. 
They have provided cross sections for single and double ionization which are
mostly in agreement with experiment for intermediate to high impact energies 
while discrepancies still persist for low energies. 
%However, some
%remaining issues have still to be clarified for the lowest impact energies. 

Antiproton collisions with molecules have been studied experimentally
in a similar way as atoms concentrating mainly on ionization cross sections
\cite{anti:hvel94} and stopping powers \cite{anti:lodi02} where for the latter
rather diverse results were obtained at low energies.
%the different results for the latter are rather divers at low energies. 
In contrast, the theoretical work on collisions involving
molecules is still comparably sparse (cf.\ \cite{anti:luhr08a} and Refs.\
therein). 
% Although antiproton collisions with molecules have been studied experimentally
% in the same way as atoms the theoretical work on collisions involving
% molecules is still comparably sparse (cf.\
% \cite{anti:hvel94,anti:luhr08a} and Refs.\ therein).
Certainly, the description of a four-particle system like a hydrogen {\em
  molecule}, consisting of two electrons and {\em two} nuclei, 
is a further step in complexity compared to a helium {\em atom}. 
%being only a three-body problem. 
%
Recently, the ionization and excitation 
cross sections \cite{anti:luhr08a,anti:luhr09b} as well as the stopping power
\cite{anti:luhr09c} for antiproton impact on molecular hydrogen were
calculated using spherical one-electron models for the hydrogen
molecule  \cite{anti:luhr08b}. They could mostly reproduce the experimental
antiproton results 
for impact energies $E \ge 90$ keV. The findings suggest, however, that for  
lower energies molecular as well as electron-electron correlation effects are
important and have to be considered.
An earlier work by Ermolaev \cite{anti:ermo93} turned out to be unsatisfactory
for $E\le 200$ keV reproducing rather atomic than molecular hydrogen.
Furthermore, two calculations, both treating the target as a molecule, 
%for molecular targets 
were performed  by Sakimoto \cite{anti:saki05} and recently by
the present authors \cite{anti:luhr09d} for molecular hydrogen ions. 
% \pb\ $+$ \htp . 
%Both studies treated the target as a molecule. 
It was shown that the calculation of only three orientations of the molecular
axis with respect to the projectile trajectory are sufficient to obtain the
ionization cross section  \cite{anti:luhr09d}.  
Currently new experimental data for antiproton collisions
with molecular hydrogen are produced using the AD facility at CERN \cite{anti:knud09a}.

In response to the renewed experimental activity and the limited theoretical 
understanding 
a full two-electron 
close-coupling method has been developed. 
%calculations for single ionization and excitation of molecular hydrogen
%from antiproton impact are presented.
Converged cross sections for single ionization and excitation of molecular
hydrogen are provided over a wide energy range from 1 keV to 6.5 MeV on a
dense energy grid. 
They demonstrate 
the importance of a full two-electron description and of the molecular
geometry including different orientations of the molecular axis as well as
the differences between atomic and molecular targets. 
To the best of the authors' knowledge no two-electron description for
antiproton impacts on molecular targets has been introduced before in this
energy range.

% \pb\ + \hmol\ data for
% ionization cross sections  \cite{anti:ande90a,anti:hvel94} and for stopping
% power \cite{anti:adam93,anti:agne95}  were measured. 
% As for the double ionization of helium targets, a considerable difference in
% the production of H$^+$ ions was observed
% between \pb\ and $p$ \cite{sct:shah82,sct:shah89} impacts. 

%Theoretically,
%however, only little has been investigated for \pb\ impacts on molecular
%targets. Very 

%%%%%%%%%%%%%%%%%%%%%%%%%%%%%%%%%%%%%%%%%%%%%%%%%%%%%%%%%%%%%%%%%% 
% 
%\section{Method} 
%\label{sec:method} 
%

%
%%%%%%%%
%\subsection{Target description}
%
%\label{sec:method_target}
%

% In this work, the previously used time-dependent close-coupling method
% \cite{anti:luhr08,anti:luhr09d} is extended to two-electron targets and
% applied to treat single ionization and excitation of the hydrogen molecule by
% antiproton impact. 

% In order to facilitate the increase of complexity of molecules
% concepts like the Born-Oppenheimer approximation have been frequently used
% and they were shown to be valuable within their range of applicability. 

The collision process is considered in a non-relativistic semi-classical way
using the impact parameter method (cf.\  \cite{sct:bran92}) which
is known to be highly accurate for impact energies $E\gtrsim 1$ keV
\cite{anti:igar04}. The quantum-mechanically treated electrons are exposed to
the Coulomb potential of the molecular nuclei as well as the heavy
projectile. The latter 
is assumed to move on a straight classical trajectory ${\bf R}(t)={\bf b} +
{\bf v} t$ given by the impact parameter ${\bf b}$ and its velocity ${\bf v}$
while $t$ is the time. 
%The {\em space-fixed} coordinate system is defined with the $x$ and $z$ axis
%being parallel to ${\bf b}$ and ${\bf v}$, respectively.

In the Born-Oppenheimer approximation the total wave function of the hydrogen
molecule separates into the product (atomic units are used
unless stated otherwise) 
\begin{equation}
  \label{eq:born_oppenheimer_hmol}
  \tilde{\psi}_k^{(\Omega)}({\bf r}_1,{\bf r}_2,{\bf \Rn}) 
             = \frac{\chi_{\nu j}^{(k)}(\Rn)}{\Rn} Y_j^m(\Theta,\Phi)
                                            \psi_k({\bf r}_1,{\bf r}_2;\Rn)\, ,
\end{equation}
where $\chi_{\nu j}^{(k)}$ are the eigenfunctions of the molecular vibration,
$Y_j^m$ the spherical harmonics, and $\Omega$\,$\equiv$\,$(\nu,j,m)$ denotes the
vibrational and rotational quantum numbers. ${\bf \Rn}=(\Rn,\Theta,\Phi)$ and
${\bf r}_{1,2}$ are the position vectors of the nuclei and the electrons,
respectively. The wave function $\psi_k({\bf r}_1,{\bf r}_2; \Rn)$  
%$\psi_{NM\Pi}({\bf r})$
satisfies the electronic part of the time-independent Schr\"odinger equation
$(\opp{H}_e)$ 
% %
% \begin{equation}
%   \label{eq:two-center-electronic-SE}
%   \opp{H}_e \, \psi^M_{N\Pi}({\bf r};\Rn) = 
%                           \epsilon^M_{N\Pi}(\Rn)\, \psi^M_{N\Pi}({\bf r};\Rn)
% \end{equation}
% %
for an unperturbed molecule at a fixed internuclear distance $\Rn$.
% , where
% $\opp{H_e}$ is the sum of the potential and the electronic part of the kinetic
% operator.
The $\psi$ are obtained together with the eigenenergies $\epsilon$ in a full
configuration-interaction (CI) calculation \cite{sfm:apal01,sfm:apal02}. The
two-electron configurations are constructed from correctly 
anti-symmetrized products of orbitals which are eigenstates of the molecular
hydrogen ion. The orbitals were obtained 
%in the same way 
as  in Ref.\ \cite{anti:luhr09d},
%using a one-center expansion of the nuclear potential. The 
where the radial part 
%of the orbitals 
is expanded in $B$ splines and the angular
part in spherical harmonics. 
More details on the extension of the close-coupling method from one-electron
\cite{anti:luhr09d,anti:luhr08} to two-electron targets are provided in
\cite{anti:luhr10a}.

%
%%%%%%%%
%\subsection{Impact parameter approximation}
%\label{sec:method_approximation} 
%

For a fixed ${\bf \Rn}$ the 
fully correlated wave function $\Psi$ of the two-electron target molecule
interacting with the antiproton is obtained by the evolution of the
time-dependent Schr\"odinger equation in real time,
%collision process can be described by the
%time-dependent Schr\"odinger equation
% 
\begin{equation} 
  \label{eq:tdSE} 
        i {\frac{\partial}{\partial t}} \Psi({\bf r}_1,{\bf r}_2,t) =  
        \left (  \opp{H}_e + \opp{V}_{\rm int}({\bf r}_1,{\bf r}_2,t)
        \right ) 
         \Psi({\bf r}_1,{\bf r}_2,t)\, , 
  \end{equation} 
where the time-dependent interaction between the electrons and the projectile
with charge $Z_p$ is expressed by 
%the time-dependent interaction potential 
% 
\begin{equation} 
  \label{eq:interaction_potential} 
         \opp{V}_{\rm int}({\bf r}_1,{\bf r}_2,t) 
         = -\frac{Z_p}{\left| {\bf r}_1-{\bf R}(t) \right|} 
           -\frac{Z_p}{\left| {\bf r}_2-{\bf R}(t) \right|} 
         %+ \frac{ Z_p Z_{\rm nuc}}{|\mathbf{R}(t)|} 
         \, . 
  \end{equation} 
% 
%The interaction of the projectile with the nuclei is not considered here
%since it would only lead to an overall phase which does not change the total
%cross sections. 
The interaction of the projectile with the nuclei, which leads only to an
overall phase, is not considered here.
%The first and the second term of the potential are due to the interaction of
%the projectile with the electron and the nuclei, respectively. 

The time-dependent scattering wave function 
\begin{equation} 
  \label{eq:psi}
  \Psi({\bf r}_1,{\bf r}_2,t)
  = \sum_k c_k(t)\,\psi_{k}({\bf r}_1,{\bf r}_2)
     \,e^{-i \epsilon_k t}
\end{equation} 
is expanded in the normalized  eigenstates $\psi_k$ of $\opp{H}_e$.
%with the eigenenergies $\epsilon_k$.
% 
%as given in Eq.\ (\ref{eq:two-center-one-center-expansion}) and $k\equiv
%NM\Pi$ stands for 
%the quantum numbers needed to label these states.
Employing this expansion in Eq.\
(\ref{eq:tdSE})  and projecting with $\psi_k$ leads to the usual set of coupled
equations for every trajectory ${\bf R}(t)$,
\begin{equation}
  \label{eq:method_coupled_equations}
  i \frac{{\rm d} c_k}{{\rm d} t} 
  = e^{i\epsilon_k t}\,
    \sum_j  c_j\,  \braket{\psi_k}{\hat{V}_{\rm int}\,|\,\psi_j} \,
           e^{-i \epsilon_j t}\, .
\end{equation}
%
%for the $c_k({\bf R}(t))$ 
%for every trajectory ${\bf R}(t)$.
The two-electron interaction matrix elements in Eq.\
(\ref{eq:method_coupled_equations}) can ---according to Eq.\
(\ref{eq:interaction_potential})--- be expressed as a sum of one-electron matrix
elements in the orbital basis.
%where the latter are calculated as explained in detail in Ref.\
%\cite{anti:luhr09d}.  
%The sum of products between the CI coefficients of $\psi_k$ and $\psi_j$ and
%the corresponding one-electron matrix elements yield
The full two-electron interaction matrix element between $\psi_k$ and
$\psi_j$ is therefore the sum of 
products between the CI coefficients of  $\psi_k$ and $\psi_j$ and 
the one-electron matrix elements 
---calculated as in \cite{anti:luhr09d}---  
between the orbitals of the corresponding CI configurations.
%which contribute to the configurations weighted
%with the product of the corresponding CI coefficients of $\psi_k$ and $\psi_j$
%\cite{anti:luhr10a}.   

% , i.e., for every impact
% parameter $b$  and every impact energy $E=(1/2)\,M_p\,v^2$,
% where $M_p$ is the projectile mass. The $c_k({\bf R}(t))$ depend of course
% also on (the fixed) ${\bf \Rn}$.
The coupled differential equations in Eq.\ (\ref{eq:method_coupled_equations})
are integrated in a finite $z$-range $-50 {\rm\ a.u.} \le z=v t \le 50$ a.u.\
with the initial conditions $c_k[{\bf R} (t_0$=$-50/v)] = \delta_{k0}$, i.e.,
the target is initially in the electronic ground state $\psi_0$ with energy
$\epsilon_0$.  
The probability for a transition into the %electronic 
final state $\psi_k$ at
$t_f=50/v$ for a fixed ${\bf \Rn} $ is given by 
\begin{equation}
  \label{eq:se_probability}
  p_k (b,E;\Rn, \Theta,\Phi) = | c_k(b,v,t_f;\Rn, \Theta,\Phi)|^2\, .
\end{equation}
In accordance with \cite{anti:saki05}, the transition probability
%
% \begin{equation}
%   \label{eq:se_probability_integrated}
%   \begin{split}
%   p_k (b,E)  &= \int \left|\chi_{\nu j}(\Rn)Y_j^m(\Theta,\Phi)\right| ^2\\
%                &\times    p_k(b,E;\Rn, \Theta,\Phi)
%                    \sin\Theta {\rm d}\Rn {\rm d}\Theta {\rm d}\Phi \, .
%                    \nonumber
% \end{split}  
% \end{equation}
%
%
\begin{eqnarray}
  \label{eq:se_probability_integrated}
  p_k (b,E)  &=& \int \left|\chi_{\nu j}(\Rn)Y_j^m(\Theta,\Phi)\right| ^2\\
             & &\times    p_k(b,E;\Rn, \Theta,\Phi)
                   \sin\Theta {\rm d}\Rn {\rm d}\Theta {\rm d}\Phi \, .
                   \nonumber
\end{eqnarray}
becomes orientation-independent by integration over ${\bf \Rn}$. The
corresponding cross section  
\begin{equation} 
  \label{eq:method_cross_section_par} 
 \sigma_{k}(E) = 2\, \pi\, \int p_{k}(b,E)\,b\; 
                         \diff{b}\,,      
\end{equation} 
can then be obtained by integration over $b$ as is done for atomic
targets which are spherical symmetric. 
The total cross sections for ionization $\sigma_{\rm ion}$ %,   
% %
% \begin{equation}
%   \label{eq:method_cross_section_ion}
%   \sigma_{\rm ion}(E) = \sum_{\epsilon_k > 0} \sigma_{k}(E) \, ,
% \end{equation}
% %
and bound-state excitation  $\sigma_{\rm exc}$ of the target %,
% %
% \begin{equation}
%   \label{eq:method_cross_section_exc}
%   \sigma_{\rm exc}(E) = \sum_{\epsilon_0 < \epsilon_k < 0} \sigma_{k}(E) \, ,
% \end{equation}
% %
can be obtained by summing up all partial cross sections $\sigma_k$ into
states $k$ [as given in Eq.\ (\ref{eq:method_cross_section_par})] with
$\epsilon_k>I_{{\rm H}_2} $ and $\epsilon_0<\epsilon_k<I_{{\rm H}_2} $,
respectively, where $I_{{\rm H}_2} $ is the first ionization potential of the
hydrogen molecule.
%all for states with
%negative energy being larger than that of the ground state $\epsilon_0$,
%respectively. 

In this work, however, $p_k $ is approximated by 
\begin{equation}
  \label{eq:method_probability_averaged}
  p_k  = \frac{1}{3} \left[
            p_k(0,0) +
            p_k\left(\frac{\pi}{2},0\right) +
            p_k\left(\frac{\pi}{2},\frac{\pi}{2}\right) \right] ,
\end{equation}
using only the probabilities $p_k(\Theta,\Phi)$ for three orthogonal
orientations of the internuclear axis with respect 
to the trajectory as is discussed in detail in Ref.\
\cite{anti:luhr09d}. It was shown that this approximation is equivalent
to Eq.\ (\ref{eq:se_probability_integrated}),
if the integration is performed with a six-point quadrature formula
\cite{sct:erre97}.  Furthermore, excellent agreement between the results using
Eqs.\ (\ref{eq:se_probability_integrated}) and
(\ref{eq:method_probability_averaged}) were obtained for antiproton
collisions with molecular hydrogen ions \cite{anti:saki05,anti:luhr09d}.

%%%%%%%%%%%%%%%%%%%%%%%
%
%\subsection{Franck-Condon approximation}
%
%\label{sec:method_Franck-Condon}
%

The dependence of the ionization cross section on the internuclear distance
$\Rn$ in antiproton collisions with the molecular hydrogen ion and the
hydrogen molecule 
%\pb\ + \htp\ 
was examined in \cite{anti:saki05} and \cite{anti:luhr08a}, respectively.
% for the range 1.5 a.u.\ $\le \Rn \le$ 3 a.u.\ in which the
% radial distribution $|\chi_{\nu j}|^2$ of the vibrational ground state
% $\chi_{00}$ is non-negligible. It has been shown that the dependence of the
% cross sections on $\Rn$ is approximately linear. A similar dependence of the
% cross sections on $\Rn$ was also obtained in calculations for  \pb\ + \hmol\
% in 
% \cite{anti:luhr08a}.  
% %
% Under the assumption that $|\chi_{\nu j}|^2$  is an even function of $\Rn -
% \bar{R}_{\rm nuc}$ and  $\sigma(\Rn)$ is linear in $\Rn$ around
% $\bar{R}_{\rm  nuc}$ the Franck-Condon (FC) approximation becomes 
% accurate as discussed, e.g., in \cite{nu:saen97b}, where $\bar{R}_{\rm
%  nuc}\equiv\langle \Rn \rangle$ is the expectation value of $\Rn$.
%
%Consequently, in \cite{anti:saki05} the  FC results 
In both cases an approximately linear dependence on $\Rn$ around the
expectation value $\langle \Rn \rangle$ of the ground state was observed
leading to Franck-Condon results which were found to be very
close to the exact cross sections obtained by an integration over
$\Rn$.
% like in Eq.\ (\ref{eq:se_probability_integrated}).  
%
%In what follows the FC approximation is used, i.e., 
In this work, 
the calculations 
%of the ionization and excitation cross sections 
are accordingly performed for %$\Rn=1.4$
$\langle \Rn \rangle = 1.4478$ 
a.u.
%being the ground state expect
%which is the
%equilibrium distance %expectation value for 
%in the ground state.
%
The two-electron basis consists of 3080 singlet states with total
azimuthal quantum numbers $M=0,\ldots,4$ and a maximum energy of $\epsilon
\approx 25$ a.u.

%%%%%%%%%%%%%%%%%%%%%%%%%%%%%%%%%%%%%%%%%%%%%%%%%%%%%%%%% 
% 
%\section{Results} 
% 
%\label{sec:results} 
% 
% 
% graphs: - stopping power for H, H2, He (3 graphs)
%         - comparison of H2 for R=1.4, 1.448, and 1.5 with lit (Discussion?)
%
%
%
%
%
%%%%%%%%%%%%%%%%%%%%%%%%%%%%%%%%%%%%%%% 
% 
% 

%
%%%%%%%%%%%%%%%%%%%%%%%%%%%%%%%%%%%%%%% 
% 
% 
%\subsection{Total cross sections } 
%\label{sec:total} 

% 
\begin{figure}[t]
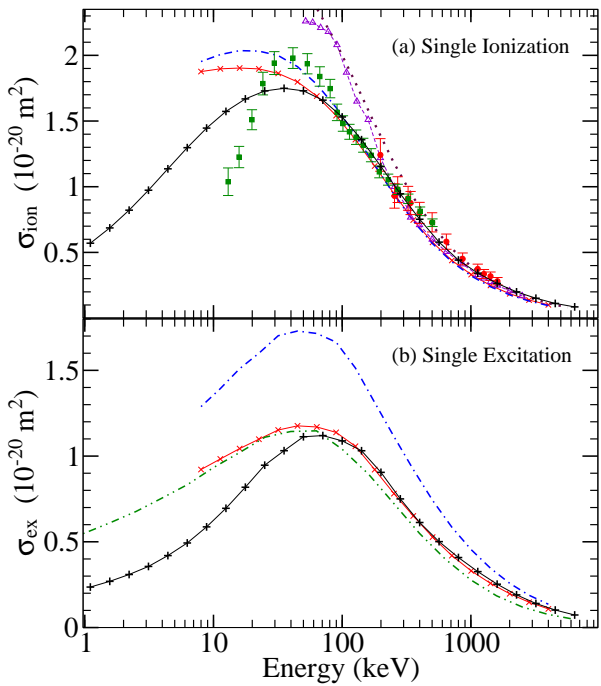
 
    \begin{center}
      \includegraphics[width=0.44\textwidth]{Pub_cs_H2_2e_lit_ION_R1p448}

      \vspace{-0.08cm}
      \includegraphics[width=0.44\textwidth]{Pub_cs_H2_2e_lit_EXC_R1p448}
      \caption{(Color online)  Cross sections for (a) single ionization
        $\sigma_{\rm ion}$ and (b) excitation $\sigma_{\rm exc}$ by antiproton 
        impact.
%        for \pb+\hmol\ as a function of the antiproton impact energy.
        Black solid curve with pluses, present results for molecular hydrogen;
        red solid curve with {\sc x}, \hmol\ model \cite{anti:luhr08a};
        blue dashed--dotted curve,  scaled hydrogen atom \hscal\
        \cite{anti:luhr08a};  
        green doubly-dotted--dashed curve, hydrogen atom \cite{anti:luhr08};
        maroon dotted curve, two times hydrogen atom \cite{anti:luhr08a};
        violet dashed curve with triangles,  \hscal\ by Ermolaev
        \cite{anti:ermo93};  
        green squares, CERN 94 \cite{anti:hvel94};
        red circles, CERN 90 \cite{anti:ande90a}.
        \label{fig:cs_H2_ion_exc} } 
    \end{center} 
\end{figure} 
% 

%
%%%%%%%%
%\subsubsection{Ionization} 
%
In Fig.\ \ref{fig:cs_H2_ion_exc}, the results of the time-dependent
close-coupling calculations for (a) single ionization and (b)
bound-state excitation of the hydrogen molecule by antiproton impact are
presented. The present results are compared to the data available in literature.
The single-ionization cross section in the top panel is in excellent agreement
with the experimental measurements for energies above 85 keV except for the
data points at 500 keV. Below 85 keV the experimental data show a small
discontinuous step and increase to a higher maximum than the present results
which is situated in both cases around 40 keV. 
Note, in the extensive convergence studies performed in this work an
enlargement of the basis always led to smaller values of the maximum.  
Below the maximum the experimental data fall off steeply in a similar way
as the data for helium which were measured in the same occasion
\cite{anti:hvel94}.   
For helium, however, the two lowest energy data points were withdrawn after a
recent remeasurement \cite{anti:knud08}. 
The currently produced %also new 
experimental hydrogen molecule data for low impact
energies may help to clarify the trend below the maximum. 

The literature results obtained 
%of the calculations 
using a model potential and a hydrogen
atom with scaled nuclear charge $Z=1.09$
\cite{anti:luhr08a,anti:luhr08b,anti:luhr09b} are able to approximate 
%mostly confirmed 
the present calculations for energies above 50 and 100 keV, respectively.
Though, they are throughout lower than the latter for these energies. 
For lower energies the models yield evidently too large cross
sections and in both cases show rather an atomic than a molecular slope by
what they reveal their atomic nature. 
Below the maximum also the lack of electron-electron correlation effects can be
expected to become severe as in the case of the helium atom
\cite{anti:knud08,anti:luhr10a}.  
The calculations by Ermolaev \cite{anti:ermo93}, using also a scaled hydrogen
atom, 
%model. His results 
are not satisfactory,
%conclusive 
since they
%close to the other models at high energies but 
follow for intermediate energies rather the data for a hydrogen atom
multiplied by a factor 2.

%\subsubsection{Excitation}

The lower panel of Fig.\ \ref{fig:cs_H2_ion_exc} compares the present
excitation cross sections for molecular hydrogen with
the existing literature, i.e., 
% the results of 
the two already mentioned models 
% potential as well as the scaled and unscaled hydrogen atom 
\cite{anti:luhr08a,anti:luhr08}. 
Obviously, the scaled hydrogen atom is not capable of reproducing the
excitation cross section for molecular hydrogen despite its reasonable
results for ionization for $E>100$ keV. 
The model potential, on the other hand, is again an excellent 
approximation for energies above 50 keV. This might have been expected since
the bound state energies and oscillator strengths of the model potential were
found to be in good agreement with those of the hydrogen molecule in contrast
to the ones predicted by the scaled hydrogen atom \cite{anti:luhr08b}.
Note, for energies above the maximum the cross section for excitation of
the hydrogen molecule is quite similar to that of atomic hydrogen while for
ionization it is rather comparable to {\em twice} the cross section of the
hydrogen atom.

\begin{figure}[t] 
    \begin{center} 
      \includegraphics[width=0.44\textwidth]{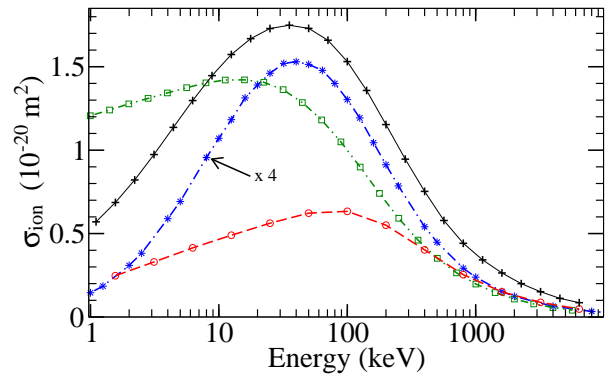} 
      \caption{(Color online) 
        Comparison of single-ionization cross sections for antiproton impact on
        different targets.
        Black solid line with pluses, present results for molecular hydrogen;
        red dashed curve with circles, helium atom \cite{anti:luhr10a};
        blue dot--dashed curve with stars, molecular hydrogen ion
        \cite{anti:luhr09d} (times 4)% scaled by 4
        ;
        green doubly-dot--dashed curve with squares, hydrogen atom
        \cite{anti:luhr08}. 
        \label{fig:cs_H2_other} } 
    \end{center} 
\end{figure} 

In Fig.\ \ref{fig:cs_H2_other}, the single-ionization cross section of
molecular hydrogen by antiproton collision is compared to results for the helium
\cite{anti:luhr10a} and hydrogen
atom \cite{anti:luhr08} and the molecular hydrogen ion \cite{anti:luhr09d}, 
where the latter is scaled by a factor 4.  
% The comparison shows that the two curves for the molecular hydrogen ion and
% the hydrogen molecule are qualitatively similar while the behavior
% of the curve for helium is more related to that of atomic hydrogen. That is,
% especially for energies below the maximum the qualitative shape of the
% ionization cross section is different for atomic and molecular targets. In the
% latter case, the cross section decreases much faster with decreasing impact
% energy than for atoms. 
The comparison shows that below the maximum the curve for molecular hydrogen
decreases much faster with decreasing energy than is the case for the
hydrogen and helium {\em atom}.
The  molecular hydrogen curve is on the other hand qualitatively similar to
that of the {\em molecular} ion. 
That is, especially for these energies the qualitative shape of
the ionization cross section seems to be different for atomic and molecular
targets. 
At low energies ionization occurs mainly in a small region close to the nuclei
where the electronic density and the expectation value of the electron
velocity are high. 
In a close encounter of the antiproton on a molecular target the electron
cloud might be more efficiently moved away from the projectile towards the
other nuclei since there is in contrast to atoms  always one positive particle
which is not neutralized by the antiproton.
 
% This might be due to the fact that for the molecules, there is always one
% positive particle which is not neutralized by the antiproton
% \cite{anti:knud09a}. 

% In the latter case, the cross section decreases much faster with decreasing impact
% energy than for atoms. 

% But already now we can say that it seems that the cross section for
% producing H2+ decreases with decreasing energy much faster than are the
% cases for H and He (see attached figure, which is for your eyes only). Maybe
% this is due to the fact the for the molecule, there is always one positive
% particle which is not neutralized by the antiproton? I look forward to see
% if your results show the same.

At high energies the single-ionization cross section for helium and 4 times
the molecular hydrogen ion are both similar to the curve of the hydrogen atom
while the results for molecular hydrogen are in good agreement with 
{\em twice} the curve for the hydrogen atom [cf.\ Fig.\
\ref{fig:cs_H2_ion_exc}(a)].  
For these energies distant encounters are dominating the ionization process.  
Accordingly, details of the targets like the exact distribution of the
positive charges become less important and the cross 
sections are mostly determined by the ionization potential.

%%%%%%%%%%%%%%%%%%%%%%%%%%%%%%%%%%%%%%% 
% 
% 
%\subsubsection{Dependence on the molecular orientation } 
%\label{sec:3directions} 
% 
% 

% 
\begin{figure}[t] 
    \begin{center} 
%       \includegraphics[width=4.642cm%0.257\textwidth
%       ]{Pub_cs_H2_3dir_R1p448_ION}%
%       \hspace{-0.075cm}%
%       \includegraphics[width=3.8cm%0.2105\textwidth
%       ]{Pub_cs_H2_3dir_R1p448_EXC} 
      \includegraphics[width=0.47\textwidth]%
      {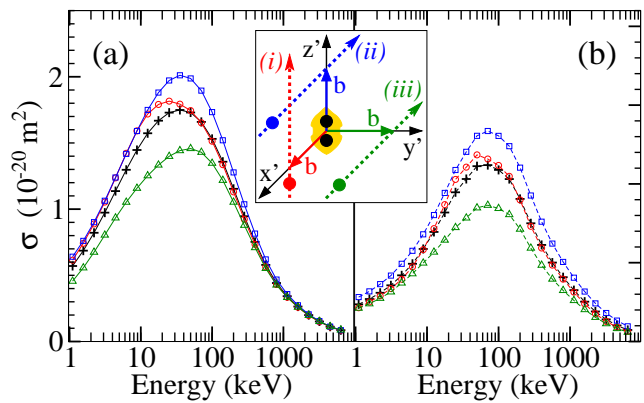}
      \caption{(Color online) Cross sections for (a) single ionization and
        (b) excitation of molecular hydrogen by antiproton collisions for
        different molecular orientations.  
        Black pluses, orientationally averaged;
        red circles, ($i$); %$(\Theta,\Phi)=(0,0)$;
        blue squares, ($ii$); % $(\pi/2,0)$;
        green triangles, ($iii$). %$(\pi/2,\pi/2)$.
%        The inset shows the high-energy tail magnified on a
%        doubly-logarithmic scale. 
        The inset shows a sketch of the three orientations in the molecule-fixed frame.
        \label{fig:cs_ion_exc_3dir}} 
    \end{center} 
\end{figure} 

In Fig.\ \ref{fig:cs_ion_exc_3dir}, the dependence on the orientation of the
molecular axis with respect to the antiproton trajectory is presented. The
cross sections  $\sigma(\Theta,\Phi)$ for (a) ionization and (b) excitation
are given for the three orthogonal orientations $(\Theta,\Phi)=$ ($i$)
$(0,0)$, ($ii$) $(\pi/2,0)$, and ($iii$) $(\pi/2,\pi/2)$ (cf.\ the sketch in
Fig.\ \ref{fig:cs_ion_exc_3dir} and Ref.\ \cite{anti:luhr09d}) revealing the
following results. First, the curves for the three different orientations
generally differ considerably especially around the maximum. 
Second, the calculation of only the parallel orientation $(i)$ reproduces for
energies above the maximum the orientationally-averaged results with less than
3\,\% relative deviation. 
Third, for lower energies at which close collisions become dominant the
consideration of the molecular geometry is inevitable.
In contrast to the findings for the molecular hydrogen ion \cite{anti:luhr09d},
the curves of orientation  $(i)$ are close to those of  $(ii)$ for ionization
below the maximum and of $(iii)$ for excitation below 5 keV.
%
% Figure \ref{fig:cs_ion_exc_3dir} shows the cross sections $\sigma
% (\Theta,\Phi)$ for the three orthogonal orientations $(\Theta,\Phi)=$ ($i$)
% $(0,0)$, ($ii$) $(\pi/2,0)$, and ($iii$) $(\pi/2,\pi/2)$ 
% for single ionization and excitation
% %which are defined as in \cite{anti:luhr09d} 
% together with the orientationally-averaged results.
% The three orientations ($i$), ($ii$), and ($iii$) and the definition of the
% according cross sections were discussed in detail for the molecular hydrogen
% ion in Ref.\ \cite{anti:luhr09d}.
% %
% Like in \cite{anti:luhr09d}, the present cross sections
% for orientation ($i$) coincides with 
% %those of 
% $\sigma_{\rm ion}$ and $\sigma_{\rm exc}$ for $E\ge 100$ keV.
% %
% For low energies, the curves of orientation ($i$) are close to those of ($ii$)
% and ($iii$) for ionization and excitation, respectively. This is in contrast
% to the findings for the molecular hydrogen ion, where for low energies the
% curves for orientations ($ii$) and ($iii$) were practically the same and the
% curves for ($i$) were clearly higher.
%
%
In general, the differences among the cross sections  
%qualitative shapes of the curves 
for the three orientations are less pronounced than for the
molecular hydrogen ion. This might be due to the 
smaller internuclear distance and the two electrons of the hydrogen molecule
making it a more spherical target.

%%%%%%%%%%%%%%%%%%%%%%%%%%%%%%%%%%%%%%% 

%%%%%%%%%%%%%%%%%%%%%%%%%%%%%%%%%%%%%%% 
%
%
% 
%\section{Summary and conclusion} 
%\label{sec:conclusion} 
% 

In conclusion, theoretical 
%benchmark 
data are presented for single ionization and excitation of molecular
hydrogen by antiproton impact for a wide energy range 
obtained with a two-electron time-dependent close-coupling method. 
%The two-electron target description is based
%on a CI calculation using products of eigenstates of the molecular hydrogen
%ion as configurations. 
%The radial part of the orbitals is expanded in $B$ splines. 
%
The experimental data are in good agreement with the present calculations at
high energies but are larger around the maximum. For energies below the
maximum the ionization cross section decreases with decreasing energy much
faster than in the cases for the hydrogen  and helium atom but in a similar
way as for the molecular hydrogen ion revealing the differences between atoms
and molecules. 
Furthermore, the importance of the molecular geometry and a full two-electron
description is demonstrated.
The present work should motivate new experimental efforts for molecular targets
at low impact energies to confirm and further extend the gained insight.
Additionally, it provides benchmark 
data for molecular collisions in general and for single 
ionization and excitation of molecular hydrogen by antiproton impact, in
particular, which might be useful for the development of molecular modes. 

The method should be further exploited to extract also quantities like, e.g.\ 
electron-energy spectra or differential excitation cross sections. Such
results may allow, especially at low energies, for the elimination of the
diversity of results for the stopping power obtained in different experiments \cite{anti:lodi02,anti:luhr09c}.

%
%\begin{center} 
% {\bf ACKNOWLEDGMENTS}  
%\end{center} 
% 

The authors gratefully acknowledge stimulating correspondence with H.\
Knudsen. This work was supported by the BMBF (FLAIR Horizon), the {\it
  Stifterverband f\"ur die deutsche Wissenschaft}, and the {\it  Fonds der
  Chemischen Industrie}.

%%%%%%%%%%%%%%%%%%%%%%%%%%%%%%%%%%%%%%%%%%%%%%%%%%%%%%%%%%%%%% 

%\bibliography{aies,anti,bsp,dia,nu,sfm,sct} 

\end{document}